\documentclass[10pt]{article}
\usepackage{graphics,epsfig}
\font\num=msbm10
\newcommand{\EE}{\hbox{\num{E}}}

\newtheorem{theorem}{Theorem}
\font\smallroman=cmr10 at 8pt
\begin{document}
\title{The Violation of Bell Inequalities in the Macroworld\footnote{To
appear in Foundations of
Physics, volume {\bf 30}, issue 10, 2000}}
\author{Diederik Aerts, Sven Aerts, Jan Broekaert and Liane Gabora}
\date{}
\maketitle
\centerline {Center Leo Apostel, Brussels Free University}
\centerline {Krijgskundestraat 33, 1160 Brussels, Belgium.}
\centerline{diraerts@vub.ac.be, saerts@vub.ac.be}
\centerline{jbroekae@vub.ac.be, lgabora@vub.ac.be}
\begin{abstract}
\noindent
We show that Bell inequalities can be violated in the macroscopic world.
The macroworld violation is illustrated using an example involving
connected vessels
of water.  We show that whether the
violation of inequalities occurs in the microworld or in the macroworld, it
is the
identification of nonidentical events that plays a crucial role.
Specifically, we
prove that if nonidentical events are consistently differentiated,
Bell-type Pitowsky inequalities are no longer
violated, even for Bohm's example of two entangled spin 1/2 quantum
particles. We show how Bell inequalities can be
violated in cognition, specifically in the
relationship between abstract concepts and specific instances of these
concepts. This supports the hypothesis that
genuine quantum structure exists in the mind. We introduce a model where
the amount of nonlocality and the degree of
quantum uncertainty are parameterized, and demonstrate that increasing
nonlocality increases the degree of violation, while
increasing quantum uncertainty decreases the degree of violation.
\end{abstract}
\bigskip
\noindent
{\it Dedication:} Marisa always stimulated interdisciplinary research
connected to quantum
mechanics, and more specifically she is very enthusiastic to the approach
that we are developing in CLEA on quantum structure in cognition. It is
therefore a pleasure for us to dedicate this paper, and particularly
the part on cognition, to her for her 60 th birthday.
\section{Introduction}
This article investigates the violation of
Bell inequalities in macroscopic
situations and analyses how this indicates the presence of
genuine quantum structure. We explicitly
challenge the common belief that quantum structure is present only in
micro-physical reality (and macroscopic
coherent systems), and present evidence that
quantum structure can be present in the macro-physical reality. We also
give an example showing the presence of quantum
structure in the mind.

Let us begin with a brief account of the most relevant historical results.
In the seventies, a sequence of experiments was
carried out to test for the presence of nonlocality in the microworld
described by quantum mechanics (Clauser 1976;
Faraci at al. 1974; Freeman and Clauser 1972; Holt and Pipkin 1973; Kasday,
Ullmann and Wu 1970) culminating in
decisive experiments by Aspect and his team in Paris (Aspect, Grangier and
Roger, 1981, 1982). They were inspired by
three important theoretical results: the EPR Paradox (Einstein, Podolsky
and Rosen, 1935), Bohm's thought
experiment (Bohm, 1951), and Bell's theorem (Bell 1964).

Einstein, Podolsky, and Rosen believed to have shown that quantum
mechanics is incomplete, in that there exist elements of reality that
cannot be described by it (Einstein, Podolsky
and Rosen, 1935; Aerts 1984, 2000). Bohm took their insight further with a
simple example: the `coupled
spin-${1\over 2}$ entity' consisting of two
particles with spin
${1\over 2}$, of which the spins are coupled such that the quantum spin
vector is a nonproduct vector representing a
singlet spin state (Bohm 1951). It was Bohm's example that inspired Bell to
formulate a condition that would test
experimentally for incompleteness. The result of his efforts are the
infamous Bell inequalities (Bell 1964). The fact
that Bell took the EPR result literally is evident from the abstract of his
1964 paper:

\begin{quotation}
\noindent
``The paradox
of Einstein, Podolsky and Rosen was advanced as an argument that quantum
theory could not be a complete theory but
should be supplemented by additional variables. These additional variables
were to restore to the theory causality and
locality. In this note that idea will be formulated mathematically and
shown to be incompatible with the statistical
predictions of quantum mechanics. It is the requirement of locality, or
more precisely that the result of a measurement
on one system be unaffected by operations on a distant system with which is
has interacted in the past, that
creates the essential difficulty."
\end{quotation}

\noindent
Bell's theorem states that statistical results of experiments performed
on a certain physical entity satisfy his inequalities if and only if the
reality in which this physical
entity is embedded is local. He believed that if experiments were performed
to test  for the presence of nonlocality as predicted
by quantum mechanics, they would show quantum mechanics to be wrong, and
locality to hold. Therefore, he
believed that he had discovered a way of showing
experimentally that quantum mechanics is
wrong. The physics community awaited the outcome of
these experiments. Today, as we know, all
of them agreed with quantum predictions, and as consequence, it is
commonly accepted that the micro-physical world is incompatible with local
realism.

One of the present authors, studying Bell inequalities from a different
perspective, developed a concrete example of a situation involving
macroscopic `classical' entities that violates
Bell  inequalities (Aerts 1981, 1982, 1985a,b). This example makes
it possible to more fully understand the origin of the violation of the
inequalities, as well as in what sense this violation indicates the
presence of quantum structure.

\section{Bell Inequalities and Clauser Horne Inequalities}
In this section we review Bell inequalities, as well as Clauser and
Horne inequalities. We first consider Bohm's original example that violates
these
inequalities in the microworld.
Finally we we put forth an example that violates them in the macroworld.

\subsection{Introduction of the Inequalities}

Bell inequalities are defined with the following experimental situation
in mind. We consider a physical
entity $S$, and four experiments $e_1$, $e_2$, $e_3$, and $e_4$ that can be
performed on the physical entity
$S$. Each of the experiments $e_i, i \in \{1, 2, 3, 4\}$ has two possible
outcomes,
respectively denoted $o_i(up)$ and $o_i(down)$. Some of the experiments
can be performed together, which
in principle leads to `coincidence' experiments $e_{ij}, i, j
\in \{1, 2, 3, 4\}$. For example
$e_i$ and $e_j$ together will be denoted $e_{ij}$. Such a coincidence
experiment $e_{ij}$ has four possible outcomes, namely
$(o_i(up), o_j(up))$, $(o_i(up), o_j(down))$, $(o_i(down), o_j(up))$ and
$(o_i(down), o_j(down))$.
Following Bell, we introduce the
expectation values $\EE_{ij},  i, j \in \{1, 2, 3, 4\}$
for these coincidence experiments, as
\begin{equation}
\begin{array}{ll}
\EE_{ij} = &  +1.P(o_i(up), o_j(up))+1.P(o_i(down), o_j(down))   \\
         & -1.P(o_i(up), o_j(down))-1.P(o_i(down), o_j(up))
\end{array}
\end{equation}
>From the assumption that the outcomes are either +1 or -1, and that the
correlation
$\EE_{ij}$ can be written as an integral over some hidden variable of a
product of the two local outcome assignments, one derives Bell inequalities:

\begin{equation} \label{bellineq} |\EE_{13} -\EE_{14}|+|\EE_{23}+ \EE_{24}|
\leq 2
\end{equation}
When Bell introduced the inequalities, he had in mind the
quantum mechanical situation originally introduced by Bohm (Bohm 1951) of
correlated spin-${1\over 2}$ particles in the singlet spin
state. Here $e_1$ and $e_2$ refer to measurements of spin at the left
location in spin directions ${\bf a_1}$ and ${\bf a_2}$, and $e_3$ and
$e_4$ refer to measurements of spin at the right location in spin
directions ${\bf a_3}$ and ${\bf a_4}$. The quantum theoretical calculation
in this situation, for well
chosen directions of spin, gives the value $2 \sqrt{2}$ for the left
member of equation (\ref{bellineq}), and
hence violates the inequalities. Since
Bell showed that the inequalities are never violated if locality holds for
the considered experimental
situation, this indicates that quantum mechanics predicts nonlocal effects
to exist (Bell 1964).

We should mention that Clauser and Horne derived other inequalities, and it
is the Clauser Horne inequalities that
have been tested experimentally (Clauser and Horne 1976). Clauser and Horne
consider the
same experimental situation as that considered by Bell. Hence we have the
coincidence
experiments $e_{13}$, $e_{14}$, $e_{23}$ and
$e_{24}$, but instead of concentrating on the expectation values
they introduce the coincidence probabilities $p_{13}$, $p_{14}$, $p_{23}$
and $p_{24}$, together with the
probabilities $p_2$ and $p_4$. Concretely, $p_{ij}$ means the probability
that the coincidence experiment
$e_{ij}$ gives the outcome
$(o_i(up), o_j(up))$, while
$p_i$ means the probability that the experiment $e_i$ gives the outcome
$o_i(up)$. The Clauser Horne inequalities then read:

\begin{equation} \label{bellineq} -1 \le p_{14} - p_{13} + p_{23} + p_{24}
- p_2 - p_4 \le 0
\end{equation}
Although the Clauser Horne
inequalities are thought to be equivalent
to Bell inequalities, they are of a slightly more general theoretical nature,
and lend themselves to Pitowsky's
generalization, which will play an important role in our theoretical
analysis.

\subsection{The `Entangled Spins ${1\over 2}$' Example}
Let us briefly consider Bohm's original example. Our physical entity $S$ is
now a pair of quantum particles of
spin-${1\over 2}$ that `fly to the left and the right' along a certain
direction $v$ of space respectively, and are
prepared in a singlet
state $\Psi_S$ for the spin (see Fig. 1). We consider four experiments
$e_1, e_2, e_3, e_4$, that
are measurements of the spin of the particles in directions ${\bf a_1, a_2,
a_3, a_4}$, that are four directions
of space orthogonal to the direction $v$ of flight of the particles. We
choose the experiments such that $e_1$ and
$e_2$ are measurements of the spin of the particle flying to the left
and $e_3$ and $e_4$ of the particle flying to the right (see Fig. 1).

\hskip 1 cm \includegraphics{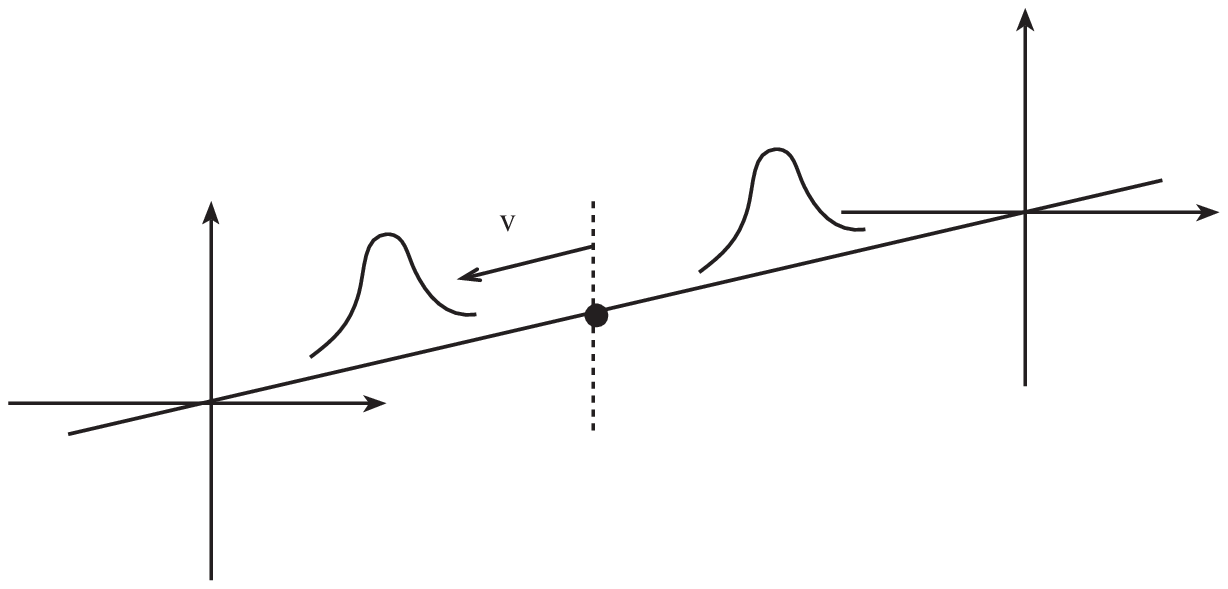}

\hskip 5 cm \noindent {\smallroman Fig.\ 1 : The singlet spin state example.}

\bigskip
\noindent
We call $p_i$ the probability that the experiment
$e_i$ gives outcome $o_i(up)$. According to quantum mechanical calculation
and considering the different
experiments, it follows:

\begin{equation} p_1 = p_2 = p_3 = p_4 = {1\over 2}
\end{equation} For the Bohm example, the experiment $e_1$ can be performed
together with the experiments
$e_3$ and $e_4$, which leads to experiment $e_{13}$ and $e_{14}$, and the
experiment
$e_2$ can also be performed together with the experiments $e_3$ and $e_4$,
which leads to experiments
$e_{23}$ and
$e_{24}$.

Quantum mechanically this corresponds to the expectation value
$\langle \sigma_1 a , \sigma_2  b \rangle=-a.b$ which gives us the well
known predictions:

\begin{equation}
\begin{array}{ll}
\EE_{13} = {\bf  - cos}\angle({\bf a_1},{\bf a_3}) & \ \   p_{13} = {1\over
2}\ {\bf
sin}^2\frac{\angle({\bf a_1},{\bf a_3})}{2}\\
\EE_{14} = {\bf  - cos}\angle({\bf a_1},{\bf a_4}) & \ \   p_{14} = {1\over
2}\ {\bf
sin}^2\frac{\angle({\bf a_1},{\bf a_4})}{2}\\
\EE_{23} = {\bf  - cos}\angle({\bf a_2},{\bf a_3}) & \ \   p_{23} = {1\over
2}\ {\bf
sin}^2\frac{\angle({\bf a_2},{\bf a_3})}{2}\\
\EE_{24} = {\bf  - cos}\angle({\bf a_2},{\bf a_4}) & \ \   p_{24} = {1\over
2}\ {\bf
sin}^2\frac{\angle({\bf a_2},{\bf a_4})}{2}
\end{array}
\end{equation} Let us first specify the situation that gives rise to a
maximal violation of Bell inequalities.
Let ${\bf a_1}, {\bf a_2}, {\bf a_3}, {\bf a_4}$ be coplanar
directions such that
$\angle({\bf a_1}, {\bf a_3}) =
\angle({\bf a_3}, {\bf a_2}) = \angle({\bf a_2}, {\bf a_4}) = 45^o$, and
$\angle({\bf a_1}, {\bf a_4}) = 135^o$ (see Fig. 2).
Then we have $\EE_{13} = \EE_{23} = \EE_{24} =
{\sqrt{2}
\over 2}$ and $\EE_{14} = - {\sqrt{2}
\over 2}$. This gives:

\begin{equation}
\begin{array}{ll} |\EE_{13} -\EE_{14}|+|\EE_{23}+ \EE_{24}| & =  |
{\sqrt{2} \over 2} -(- {\sqrt{2} \over
2}) | + | {\sqrt{2} \over 2} + {\sqrt{2} \over 2} |  \\
 & =  + 2\sqrt{2} \\ & > + 2
\end{array}
\end{equation}
which shows that Bell inequalities are violated.

\bigskip

\hskip 3.1 cm \includegraphics{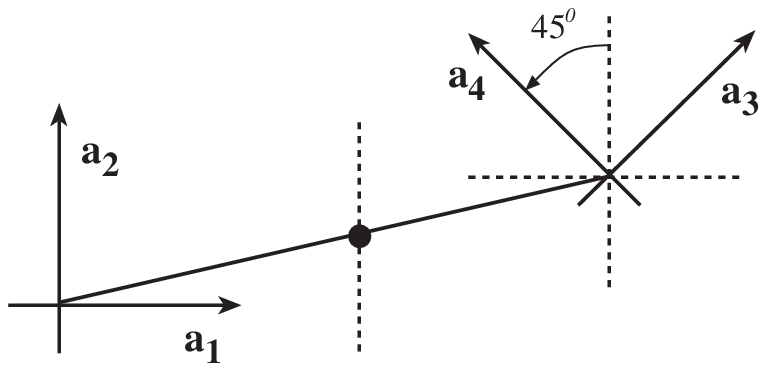}

\medskip
\centerline{\smallroman  Fig.\ 2 : The violation of Bell
inequalities by the singlet spin state example.}

\medskip
\bigskip
\noindent
To
violate the Clauser Horne inequalities,
we need to make another choice for the spin directions. Let us choose ${\bf
a_1}, {\bf a_2}, {\bf a_4}$
again coplanar with
$\angle({\bf a_1},{\bf a_2})=\angle({\bf a_1},{\bf a_4})=\angle({\bf
a_2},{\bf a_4})=120^{\circ}$ and ${\bf
a_2}={\bf a_3}$ (see  Fig. 3). The set of probabilities that we consider
for the Clauser Horne
inequalities is then given by $p_1 = {1\over 2}, p_2 = {1\over 2}, p_3 =
{1\over 2} ,p_4 = {1\over  2},
p_{12} = {3\over 8}, p_{14} = {3\over 8},p_{23} = 0,p_{24} = {3\over 8}$.
This gives:

\begin{equation}
\begin{array}{ll} p_{14} - p_{13} + p_{23} + p_{24} - p_2 - p_4 & = +1 - 0
+ 1 + 1 - 1 - 1 \\ & = + 1 \\ & >
0
\end{array}
\end{equation} which shows that also the Clauser Horne inequalities are
violated (see Fig. 3).

\bigskip

\hskip 3.1 cm \includegraphics{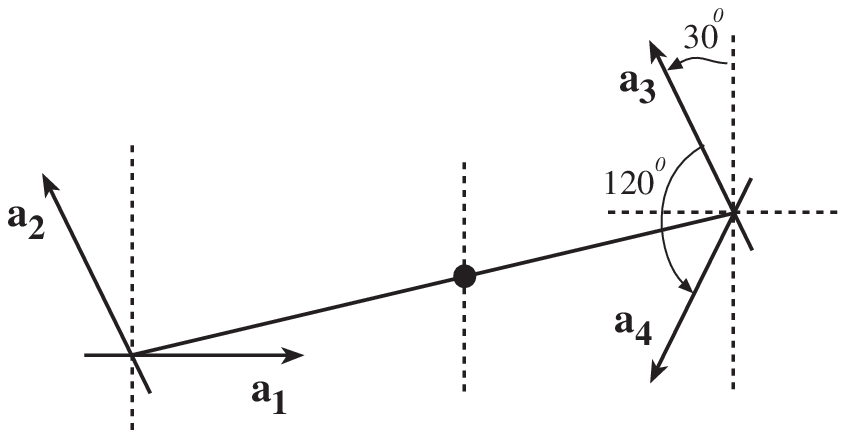}

\medskip
\centerline{\smallroman  Fig.\ 3 : The violation of Clauser Horne
inequalities by the singlet spin state
example.}
\medskip

\subsection{The `Vessels of Water' Example} \label{section03}

We now review an example of a macroscopic situation where Bell inequalities
and Clauser Horne inequalities are violated (Aerts 1981, 1982, 1985a,b).
Following this, we analyze some aspects of the example in a new way.

\hskip 4 cm \includegraphics{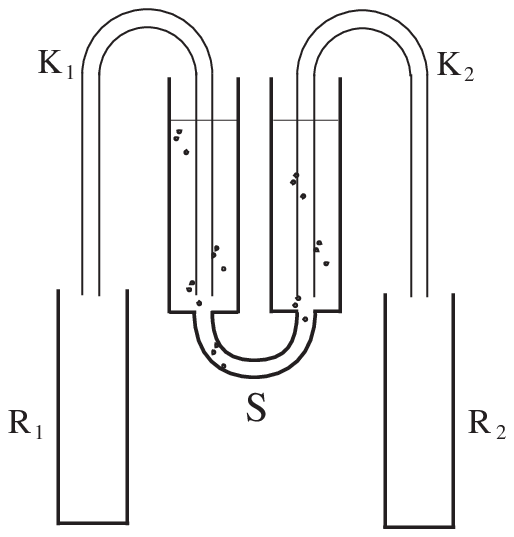}

\medskip
\noindent {\baselineskip= 7pt \smallroman  Fig.\ 4 : The vessels of water
example
violating Bell inequalities. The entity $S$
consists of two vessels containing 20 liters of water that are connected by
a tube. Experiments are performed on
both sides of the entity $S$ by introducing syphons $K_1$ and $K_2$ in the
respective vessels and pouring out the
water and collecting it in reference vessels $R_1$ and $R_2$. Carefully
chosen experiments reveal that Bell
inequalities are violated by this entity $S$. \par}

\bigskip
\noindent
Consider an entity $S$ which is a container with 20 liters of transparent water
(see Fig. 4), in a state $s$ such
that the container is placed in the gravitational field of the earth, with
its bottom horizontal. We introduce the experiment $e_1$ that
consists of putting a siphon
$K_1$ in the container of water at the left, taking out water using the
siphon, and collecting this
water in a reference vessel $R_1$ placed to the left of the
container. If we collect more than 10
liters of water, we call the outcome $o_1(up)$, and if we collect less or
equal to
10 liters, we call the
outcome $o_1(down)$. We introduce another experiment
$e_2$ that consists of taking with a little spoon, from the left, a bit of
the water, and determining whether
it is transparent. We call the outcome $o_2(up)$ when the water is
transparent and the outcome $o_2(down)$
when it is not. We introduce the experiment $e_3$ that consists of putting
a siphon
$K_3$ in the container of water at the right, taking out water using
the siphon, and collecting this
water in a reference vessel $R_3$ to the right of the
container. If we collect more or equal to 10
liters of water, we call the outcome $o_3(up)$, and if we collect less than
10 liters, we call the
outcome $o_3(down)$. We also introduce the experiment $e_4$ which is
analogous to experiment $e_2$, except
that we perform it to the right of the container.

\par Clearly, for the container of water being in state $s$, experiments
$e_1$ and $e_3$ give with
certainty the outcome $o_1(up)$ and
$o_3(up)$, which shows that $p_1 = p_3
= 1$. Experiments $e_2$ and $e_4$ give with certainty the outcome
$o_2(up)$ and $o_4(up)$, which shows that $p_2 =
p_4 = 1$.

\par The experiment $e_1$ can be performed together with experiments $e_3$
and $e_4$, and we denote the
coincidence experiments
$e_{13}$ and $e_{14}$. Also, experiment $e_2$ can be performed together
with experiments $e_3$ and
$e_4$, and we denote the coincidence experiments
$e_{23}$ and $e_{24}$. For the container in state $s$, the coincidence
experiment $e_{13}$ always gives
one of the outcomes $(o_1(up), o_3(down))$ or $(o_1(down), o_3(up))$, since
more than 10 liters of water can never come out of the vessel at both sides.
This shows that $\EE_{13} =-1$ and
$p_{13} = 0$. The coincidence experiment
$e_{14}$ always gives the outcome $(o_1(up), o_4(up))$ which shows that
$\EE_{14} = +1$ and $p_{14} = +1$, and the coincidence experiment $e_{23}$
always gives the outcome
$(o_2(up), o_3(up))$ which shows that
$\EE_{23} = +1$ and $p_{23} = +1$. Clearly experiment $e_{24}$ always
gives the outcome
$(o_2(up), o_4(up))$ which shows that $\EE_{24} = +1$ and $p_{24} = +1$.
Let us now calculate the terms of
Bell inequalities,
\begin{equation}
\begin{array}{ll} |\EE_{13} -\EE_{14}|+|\EE_{23}+ \EE_{24}| & =  | -1 -1 |
+ | +1 +1 |  \\
 & =  + 2 + 2 \\
 & = + 4
\end{array}
\end{equation}
and of the Clauser Horne inequalities,
\begin{equation}
\begin{array}{ll}
p_{14} - p_{13} + p_{23} + p_{24} - p_2 - p_4 & = +1 - 0 + 1 + 1 - 1 - 1 \\
& = + 1
\end{array}
\end{equation} This shows that
Bell inequalities and Clauser Horne inequalities can be violated in
macroscopic reality. It is even so
that the example violates the inequalities more
than the original quantum example
of the two coupled spin-${1\over 2}$ entities.
In section 5 we analyze why this is the case, and show that this sheds new
light on the underlying mechanism that leads to the violation of
the inequalities.

\section{The Inequalities and Distinguishing Events}

In the macroscopic example of the vessels of water connected by a tube, we
can see and understand why the inequalities
are violated. This is not the case for the micro-physical Bohm example of
coupled spins.
Szabo (pers. com.) suggested that the macroscopic violation of Bell
inequalities
by the vessels of water example does not
have the same `status' as the microscopic violation in the Bohm example
of entangled spins, because events are identified
that are not identical. This idea was first considered in Aerts and Szabo,
1993. Here we investigate it more carefully and
we will find that it leads to a
deeper insight into the meaning of the violation of the inequalities.

Let us state more clearly what we mean by reconsidering the vessels of
water example from section
\ref{section03}, where we let the experiments
$e_1$, $e_2$, $e_3$ and $e_4$
correspond with possible events
$A_1(up)$ and $A_1(down)$,
$A_2(up)$ and $A_2(down)$, $A_3(up)$ and $A_3(down)$ and $A_4(up)$ and
$A_4(down)$. This means that event
$A_1(up)$ is the physical event that happens when experiment
$e_1$ is carried out, and outcome $o_1(up)$ occurs. The same for the
other events. When event
$A_1(up)$ occurs together with event $A_3(down)$, hence during the
performance of the experiments $e_{12}$,
then it is definitely a different event from event $A_1(up)$ that
occurs together with event
$A_4(up)$, hence during the performance of the experiment $e_{14}$. Szabo's
idea was that this `fact'
would be at the origin of the violation of Bell inequalities for the
macroscopic vessel of water example.
In this sense the macroscopic violation would not be a genuine
violation as compared to the microscopic.

Of course, one is tempted to ask the same question in the quantum case:
is the violation in the microscopic world perhaps due to a lack of
distinguishing
events that are in fact not identical? Perhaps what is true for
the vessels of water example is also true of the Bohm example?
 Let us find out by systematically distinguishing between events at the
left (of the vessels of
water or of the entangled spins) that are made together with different
events at
the right. In this way, we get more than four events, and
unfortunately the original Bell inequalities are out of their domain of
applicability.  However Pitowsky has developed a generalization of Bell
inequalities where any number of experiments and events can be taken into
account, and as a consequence we can check whether the new situation
violates Pitowsky inequalities.
If Pitowsky inequalities would not be violated in the vessels of water model,
while for the microscopic Bohm example they would, then this would `prove'
the different
status of the two examples, the macroscopic being
`false', due to lack of correctly distinguishing between events, and the
microscopic being genuine. Let us first introduce Pitowsky inequalities
to see how the problem can be reformulated.

\subsection{Pitowsky Inequalities}

Pitowsky proved (see theorem
\ref{theorem01}) that the situation where
Bell-type inequalities are satisfied is equivalent to the situation where,
for a set of probabilities
connected to the outcomes of the considered experiments, there exists a
Kolmogorovian probability model. Or, as some may want to paraphrase it,
the probabilities are classical (Pitowsky 1989).
To put forward Pitowsky inequalities, we have to
introduce some mathematical concepts. Let
$S$ be a set of pairs of integers from $\{1, 2, ... , n\}$ that is,
\begin{equation} S\subseteq\left\{\{i,j\}\mid1\leq i< j\leq n\right\}
\end{equation} Let $R(n,S)$ denote the real space of all functions $f :
\{1, 2, ... , n\}
\cup S \mapsto${\num{R}}. We shall denote
vectors in $R(n, S)$ by ${\bf f} =
(f_1,f_2,...f_n,...f_{ij},...)$, where the $f_{ij}$ appear in a
lexicographic order on the
$i, j's$. Let $\{0, 1\}^n$ be the set of all $n$-tuples of zeroes and
one's.  We shall denote elements of
$\{0, 1\}^n$ by $\varepsilon = (\varepsilon_1, \varepsilon_2, ... ,
\varepsilon_n)$ where $\varepsilon_j \in \{0, 1\}$. For each
$\varepsilon\in\{0,1\}^n$ let
${\bf u^{\varepsilon}}$ be the following vector in
$R(n,S)$:
\begin{eqnarray} u^{\varepsilon}_j &=& \varepsilon_j \quad \quad 1\leq j
\leq n \\ u^{\varepsilon}_{ij} &=&
\varepsilon_i\varepsilon_j \quad \quad \{i,j\} \in S
\end{eqnarray} The classical correlation polytope $C(n,S)$ is the closed
convex hull in
$R(n,S)$  of all $2^n$ possible vectors ${\bf
u}^{\varepsilon}$, ${\varepsilon\in\{0,1\}^n}$:

\begin{theorem} [Pitowsky, 1989] \label{theorem01} Let ${\bf p} = (p_1, ...
, p_n, ... , p_{ij}, ... \}$ be
a vector in $R(n, S)$. Then ${\bf p} \in C(n, S)$ if there is a
Kolmogorovian probability space $(X, {\cal
M} , \mu)$ and (not necessarily distinct) events $A_1, A_2, ... , A_n \in
{\cal M}$ such that:
\begin{equation} p_i =  \mu(A_i) \quad 1 \le i \le n \quad \quad p_{ij} =
\mu(A_i \cap A_j)
\quad \{i, j\} \in S
\end{equation}
Where $X$ is ..., ${\cal M}$ is the space of events and $\mu$ the
probability measure
\end{theorem}
To illustrate the
theorem and at the same time the connection with Bell inequalities and
the Clauser Horne inequalities, let
us consider some specific examples of Pitowsky's theorem.

\medskip
\noindent The case $n=4$ and $S=\left\{ \left\{
1,3\right\},\left\{1,4\right\},\left\{ 2,3\right\},\left\{
2,4\right\}
\right\}$. The  condition ${\bf p}\in C(n,S)$ is then equivalent to the
Clauser-Horne inequalities (see
(\ref{bellineq})):
\begin{equation}
\begin{array}{cl}
0\leq p_{ij}\leq p_i \leq 1 &  \\
0\leq p_{ij}\leq p_j \leq 1  & i=1,2  \quad j=3,4 \\
p_i+p_j-p_{ij} \leq 1 & \\
-1\leq p_{13}+p_{14}+p_{24}-p_{23}-p_1-p_4 \leq 0 & \\
-1\leq p_{23}+p_{24}+p_{14}-p_{13}-p_2-p_4 \leq 0 & \\
-1\leq p_{14}+p_{13}+p_{23}-p_{24}-p_1-p_3 \leq 0 & \\
-1\leq p_{24}+p_{23}+p_{13}-p_{14}-p_2-p_3 \leq 0 &
\end{array}
\end{equation}

\medskip
\noindent The case $n=3$ and $S = \{\{1,2\},\{1,3\},\{2,3\}\}$. We find
then the following inequalities
equivalent to the condition ${\bf p}\in C(n,S)$:
\begin{equation}
\begin{array}{cl}
0 \leq p_{ij}\leq p_i \leq 1 &  \\
0 \leq p_{ij}\leq p_j \leq 1 &  1 \leq i < j \leq 3\\
p_i+p_j-p_{ij}  \leq1 &   \\
p_1+p_2+p_3-p_{12}-p_{13}-p_{23} \leq 1 \leq 0 & \\
p_1-p_{12}-p_{13}+p_{23} \leq 0 & \\
p_2-p_{12}-p_{23}+p_{13} \leq 0 & \\
p_3-p_{13}-p_{23}+p_{12} \leq 0 &
\end{array}
\end{equation} It can be shown that these inequalities are equivalent to the
original Bell inequalities
(Pitowsky 1989).

\subsection{The Genuine Quantum Mechanical Nature of the Macroscopic
Violations}
Let us now introduce a new
situation wherein the events are
systematically distinguished. For the vessels of water example, we
introduce the following
events: Event $E_1$ corresponds
to the physical process of experiment $e_1$, leading to outcome
$o_1(up)$, performed together with experiment $e_3$ leading to
outcome $o_3(up)$. Event $E_2$
corresponds to the physical process of experiment $e_1$ leading to
outcome $o_1(up)$, performed together
with experiment $e_4$, leading to outcome
$o_4(up)$. In order to introduce the other events and avoid repetition,
 we abbreviate event $E_2$ as follows:
$$ E_2= [O(e_1)=o_1(up)  \;  \& \; O(e_4)=o_4(up)]$$
The other events can then be written analogously as:
\begin{equation}
\begin{array}{ll}
E_3= [O(e_2)=o_2(up)  \;  \& \; O(e_3)=o_3(up)]   & \ \ \ E_4=
[O(e_2)=o_2(up)  \;  \& \; O(e_4)=o_4(up)] \cr
E_5= [O(e_3)=o_3(up)  \;  \& \; O(e_1)=o_1(up)]   & \ \ \ E_6=
[O(e_3)=o_3(up)  \;  \& \; O(e_2)=o_2(up)] \cr
E_7= [O(e_4)=o_4(up)  \;  \& \; O(e_1)=o_1(up)]   & \ \ \ E_8=
[O(e_4)=o_4(up)  \;  \& \; O(e_2)=o_2(up)]
\end{array}
\end{equation}
The physical process of the joint experiment $e_{13}$
corresponds then to the joint event $E_1
\wedge E_5$, the physical process of the joint experiment $e_{14}$ to the
joint event $E_2 \wedge E_7$, the
physical process of the joint experiment $e_{23}$ to the joint event $E_3
\wedge E_6$, and the physical
process of the joint experiment $e_{24}$ to the joint event $E_4
\wedge E_8$.

Having distinguished the events in this way, we are certain the different
joint experiments give rise to real joint events. We can now apply
Pitowsky's theorem to the set of events
$E_1, E_2, E_3, E_4, E_5, E_6, E_7, E_8, E_1 \wedge E_5, E_2\wedge E_7,
E_3\wedge E_6, E_4\wedge E_8$.

Suppose that there is an equal probability of experiment $e_1$ being
performed with $e_3$ or $e_4$, and similarly for the joint
performance of $e_2$ with
$e_3$ or $e_4$. According to this assumption, the observed probabilities are:
\begin{equation}
\begin{array}{c} p(E_1) =p(E_5)=0  \\ p(E_2) =p(E_3)=p(E_4) =
p(E_6)=p(E_7)=p(E_8)={{1\over  2}} \\ p(E_1
\wedge E_5)=0 \\ p(E_2 \wedge E_7)=p(E_3 \wedge E_6)=p(E_4 \wedge
E_8)={{1\over 4}}
\end{array}
\end{equation}
The obtained probability vector is then ${\bf p}=(0,{{1\over
2}},{1\over 2},{1\over
2},0,{1\over 2},{1\over 2},{1\over 2},0,{1\over 4},{1\over 4},{1\over 4})$.
Applying Pitowsky's approach, we could directly calculate
 that this probability vector is contained in
the convex hull of the
corresponding space, and hence as a consequence of Pitowsky's theorem it allows
a Kolmogorovian probability
representation (Aerts and Szabo 1993). This means that after the
distinction between events has been made, the
vessels of water example no longer violates Pitowsky inequalities.
An important question remains: would the violation of the
inequalities similarly vanish for the
microscopic spin example? Let us, exactly as we have done in the vessels of
water example, distinguish the
events we are not certain we can identify, for the case
of the correlated spin situation.
Again we find 8 events: Event $E_1$ corresponds to the
physical process of experiment $e_1$
leading to outcome $o_1(up)$, performed together with experiment $e_3$,
leading to outcome
$o_3(up)$. Analogously, events $E_2$, $E_3$, $E_3$,
$E_4$, $E_5$, $E_6$, $E_7$, $E_8$ are introduced. Again, the physical
process of the joint experiment
$e_{13}$ corresponds then to the joint event $E_1 \wedge E_5$, the physical
process of the joint experiment
$e_{14}$ to the joint event $E_2 \wedge E_7$, the physical process of the
joint experiment $e_{23}$ to the
joint event $E_3 \wedge E_6$, and the physical process of the joint
experiment $e_{24}$ to the joint event
$E_4
\wedge E_8$.

Suppose that directions ${\bf a_1}$ or ${\bf a_2}$, as well as ${\bf a_3}$
or ${\bf a_4}$, are chosen with the
same probability at both sides. According to this assumption the observed
probabilities are:

\begin{equation}
\begin{array}{c} p_i = p(E_i) = {1\over 4} \quad 1 \le i \le 8  \\ p_{15} =
p(E_1 \wedge E_5) = {1\over
4}{\bf sin^2 \angle({\bf a_1},{\bf a_3})} = {3\over 16} \\ p_{27} = p(E_2
\wedge E_7) = {1\over 4}{\bf
sin^2\angle({\bf a_1},{\bf a_4})} = {3\over 16} \\ p{36} = p(E_3 \wedge
E_6) = {1\over 4}{\bf sin^2
\angle({\bf a_2},{\bf a_3})} = 0 \\ p_{48} = p(E_4 \wedge E_8) = {1\over
4}{\bf sin^2\angle({\bf a_2},{\bf
a_4})} = {3\over 16}
\end{array}
\end{equation} The question is whether the correlation vector ${\bf p} =
({1\over 4},{1\over 4}, {1\over
4},{1\over 4},{1\over 4},{1\over 4},{1\over 4},{1\over 4},{1\over
4},{1\over 4},{3\over 16}, {3\over 16}, 0,
{3\over 16})$ admits a Kolmogorovian representation. To answer this
question, we have to check whether it is
inside the corresponding classical correlation polytope
$C(8,S)$. There are no derived inequalities for $n=8$, expressing the
condition ${\bf p} \in C(n,S)$. Lacking such inequalities we must  directly
check the geometric condition
${\bf p} \in C(n, S)$. We were able to do this for the vessels of water
example because of the simplicity of
the correlation vector, but we had no general way to do this. It is
however possible to prove the existence
of a Kolmogorovian representation in a general way:

\begin{theorem} Let events $E_1, E_2, ..., E_n$ and a set of indices $S$ be
given such that non of the
indices appears in two different elements of
$S$. Assume that for each pair $\{i,j\} \in S$ the restricted correlation
vector $p_{\{i,j\}} =
(p(E_i),p_(E_j), p(E_i \wedge E_j))$ has an $(X_{\{i,j\}}, \mu_{\{i,j\}})$
Kolmogorovian representation. Then
the product space $(X_{\{i_1,j_1\}} \times X_{\{i_2,j_2\}} \times ...
\times X_{\{i_{\vert S
\vert}, j_{\vert S \vert}\}}, \mu_{\{i_1,j_1\}} \times \mu_{\{i_2,j_2\}}
\times ... \times
\mu_{\{i_{\vert S \vert}, j_{\vert S \vert}\}})$ provides a Kolmogorovian
representation for the whole
correlation vector ${\bf p}$.
\end{theorem} This theorem shows that if the distinctions that we have
explained are made, the inequalities
corresponding to the situation will no longer be violated. This also
means that we can state that the
macroscopic violation, certainly with respect to the distinction or
identification of events, is as genuine
as the microscopic violation of the inequalities.

\section{The Violation of Bell Inequalities in Cognition}

In this section we show how Bell inequalities are violated in the mind in
virtue of the relationship between abstract
concepts and specific instances of them. We start with a thought experiment
that outlines a possible scenario wherein
this sort of violation of Bell inequalities reveals itself. This example
was first presented in Aerts and Gabora 1999. We
then briefly discuss implications for cognition.

\subsection{How Concepts Violate Bell inequalities}
To make things more concrete we begin with an example. Keynote players in
this example are the two cats, Glimmer and
Inkling, that live at our research center (Fig. 5).
The experimental situation has been set up by one of the authors (Diederik)
to show that the mind of another of the authors
(Liane) violates Bell inequalities. The situation is as follows. On the
table where Liane prepares the food for the cats
is a little note that says: `Think of one of the cats now'. To show that
Bell inequalities are violated we must
introduce four experiments $e_1, e_2, e_3$ and $e_4$. Experiment $e_1$
consists of Glimmer showing up at the instant
Liane reads the note. If, as a result of the appearance of Glimmer and
Liane reading the note, the state of her mind is
changed from the more general concept `cat' to the instance `Glimmer', we
call the outcome $o_1(up)$, and if it is
changed to the instance `Inkling', we call the outcome $o_1(down)$.
Experiment $e_3$ consists of Inkling showing up at
the instant that Liane reads the note. We call the outcome $o_3(up)$ if the
state of her mind is changed to the instance
`Inkling', and $o_3(down)$ if it is changed to the instance `Glimmer', as a
result of the appearance of Inkling and
Liane reading the note. The coincidence experiment $e_{13}$ consists of
Glimmer and Inkling both showing up when Liane
reads the note. The outcome is ($o_1(up)$, $o_3(down)$) if the state of her
mind is changed to the instance `Glimmer',
and ($o_1(down)$, $o_3(up)$) if it changes to the instance `Inkling' as a
consequence of their appearance and the
reading of the note.

\medskip
\hskip 3.2 cm \includegraphics{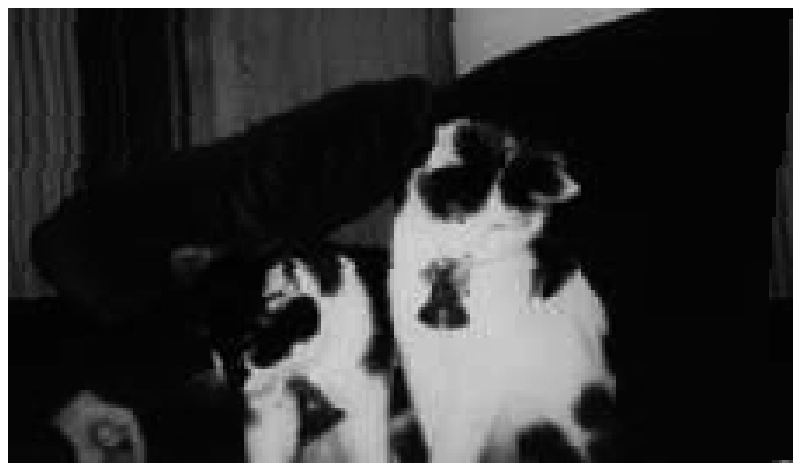}

\bigskip
\noindent
{\baselineskip= 7pt \smallroman Fig\ 5: Inkling (left) and Glimmer (right).
This picture was taken
before Glimmer decided that the quantum cat superstar life was not for him
and started to remove his
bell.\par}

\bigskip
\noindent
Now it is necessary to know that occasionally the secretary puts bells on
the cats' necks, and occasionally she takes the bells off. Thus, when Liane
comes to work, she does not know whether or not the cats will be wearing
bells, and she is always curious to know. Whenever she sees one of the
cats, she eagerly both looks and listens for the bell. Experiment $e_2$
consists of Liane seeing Inkling and noticing that she hears a bell ring or
doesn't. We give the outcome $o_2(up)$ to the experiment $e_2$ when Liane
hears the bell, and $o_2(down)$ when she does not. Experiment $e_4$ is
identical to
experiment $e_2$ except that Inkling is interchanged with Glimmer.
The coincidence experiment $e_{14}$ consists of Liane reading the note, and
Glimmer showing up, and her listening to whether a bell is ringing or not.
It has four possible outcomes: ($o_1(up)$, $o_4(up)$) when the state of
Liane's mind is changed to the instance `Glimmer' and she hears a bell;
($o_1(up)$,
$o_4(down)$) when the state of her mind is changed to the instance
`Glimmer' and she does not
hear a bell; ($o_1(down)$, $o_4(up)$) when the state of her mind is changed
to the instance `Inkling' and she hears a bell and ($o_1(down)$,
$o_4(down)$) when
the state of her mind is changed to the instance `Inkling' and she does not
hear a bell.
The coincidence experiment $e_{23}$ is defined analogously. It consists of
Liane reading the note and Inkling showing up and her listening to whether
a bell is ringing or not. It too has four possible outcomes: ($o_2(up)$,
$o_3(up)$) when she hears a bell and the state of her mind is changed to
the instance `Inkling'; ($o_2(up)$, $o_3(down)$) when she hears a bell and
the state of her mind is changed to the instance `Glimmer'; ($o_1(down)$,
$o_3(up)$) when she does not hear a bell and the state of her mind is
changed to the instance `Inkling' and ($o_1(down)$, $o_3(down)$) when she
does not hear a bell and the state of her mind is changed to the instance
`Glimmer'. The coincidence experiment $e_{24}$ is the experiment where
Glimmer and Inkling show up and Liane listens to see whether she hears the
ringing of bells. It has outcome ($o_2(up)$, $o_4(up)$) when both cats wear
bells,
($o_2(up)$, $o_4(down)$) when only Inkling wears a bell, ($o_2(down)$,
$o_4(up)$) when only Glimmer wears a bell and ($o_2(down)$, $o_4(down)$)
when neither cat wears a bell.

We now formulate the necessary conditions
such that Bell inequalities are violated in this experiment:
\begin{description}
\item[{(1)}] The categorical concept `cat' is activated in Liane's mind.
\item[{(2)}] She does what is written on the note.
\item[{(3)}]  When she sees Glimmer, there is a change of state, and the
categorical concept `cat' changes to the instance 'Glimmer',
and when she sees Inkling it changes to the instance 'Inkling'.
\item[{(4)}] Both cats are wearing bells around their necks.
\end{description}
The coincidence experiment $e_{13}$ gives outcome ($o_1(up)$, $o_3(down)$)
or ($o_1(down)$, $o_3(up)$) because indeed from (2) it follows that Liane
will think of Glimmer or Inkling. This means that $\EE_{13} = -1$. The
coincidence experiment $e_{14}$
gives outcome ($o_1(up)$, $o_4(up)$), because from (3) and (4) it follows
that she thinks of Glimmer and hears the bell. Hence $\EE_{14} = +1$. The
coincidence experiment $e_{23}$ also gives outcome ($o_2(up)$, $o_3(up)$),
because from (3) and (4) it follows that she thinks of Inkling and hears
the bell. Hence
$\EE_{23} = +1$. The coincidence experiment $e_{24}$ gives ($o_2(up)$,
$o_4(up)$), because from (4) it follows that she hears two bells. Hence
$\EE_{24} = +1$. As a consequence we have:

\begin{equation}
| \EE_{13} - \EE_{14} | + | \EE_{23} + \EE_{24} | = +4
\end{equation}
The reason that Bell inequalities are violated is that Liane's state of
mind changes from activation of the abstract
categorical concept `cat',  to activation of either `Glimmer' or `Inkling'.
We can thus view the state `cat' as an
entangled state of these two instances of it.

We end this section by saying that we apologize for the pun on Bell's
name, but it seemed like a good way to ring in these new ideas.

\subsection{The Nonlocality of Concepts}
Our example shows that concepts in the mind violate Bell inequalities, and
hence entail nonlocality in the sense that
physicists use the concept. This violation of Bell inequalities
takes place within the associative network
of concepts and episodic memories constituting an internal model of
reality, or worldview. We now briefly investigate how
this cognitive source of nonlocality arises, and its implications for
cognition and our understanding of reality.

As a first approximation, we can say that the nonlocality of stored
experiences and concepts arises from their
distributed nature. Each concept is stored in many
memory locations; likewise, each location participates in the storage of
many concepts. In order for the mind to be
capable of generating a stream of meaningfully-related yet potentially
creative remindings, the degree of this
distribution must fall within an intermediate range. Thus, a given
experience activates not just one location in memory,
nor does it activate every memory location to an equal degree, but
activation is distributed across many memory
locations, with degree of activation falling with distance from the most
activated one. Fig. 6 shows schematically how
this feature of memory is sometimes modeled in neural networks using a
radial basis function (RBF) (Hancock et al.,
1991; Holden and Niranjan, 1997; Lu et al. 1997; Willshaw and Dayan, 1990).

\bigskip
\hskip 3.2 cm \includegraphics{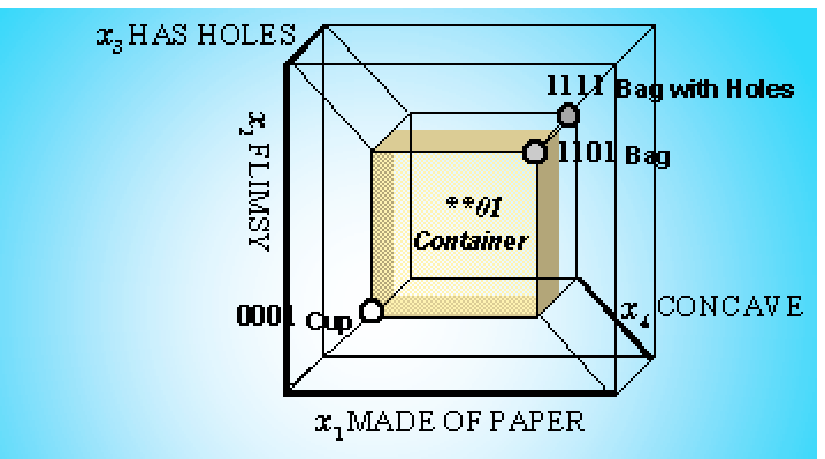}

\bigskip
\noindent
{\baselineskip= 7pt \smallroman Fig\ 6: Highly schematized diagram of a
stimulus input activating two
dimensions of a memory or conceptual space. Each vertex represents a
possible memory location, and black dots represent actual location in
memory. Activation is maximal at the center k of the RBF, and tapers off in
all directions according to a Gaussian distribution of width s.\par}

\bigskip
\noindent
Memory is also content addressable, meaning that there is a systematic
relationship between the content of an experience, and the place in memory
where it gets stored (and from which material for the next instant of
experience is sometimes evoked). Thus not only is it is not localized as an
episodic memory or conceptual entity in conceptual space, but it is also
not localized with respect to its physical storage location in the brain.

\subsection{The Relationship between Nonlocality and Degree of Abstraction}
The more abstract a concept, the greater the number of other concepts that
are expected to fall within a given distance of it in conceptual space, and
therefore be potentially evoked by it. For example, Fig. 7 shows how the
concept of `container' is less localized than the concept of `bag'. The
concept of `container' does not just activate concepts like `cup', it
derives its very existence from them. Similarly, once `cup' has been
identified as an instance of `container', it is forever after affected by
it. To activate `bag' is to instantaneously affect `container', which is to
instantaneously affects the concept `thing', from which `container' derives
it's identity, and so forth.

\bigskip
\hskip 4.5 cm \includegraphics{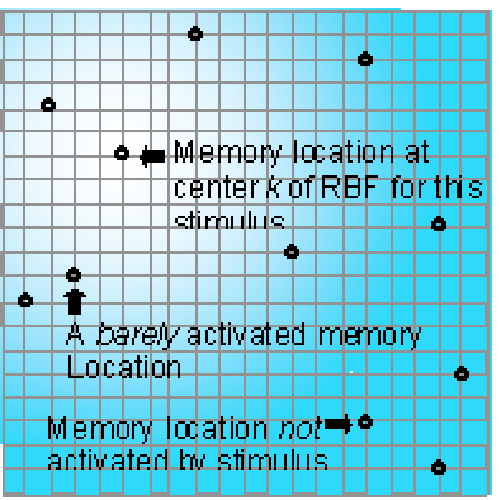}

\bigskip
\noindent
{\baselineskip= 7pt \smallroman  Fig\ 7: A four-dimensional hypercube that
schematically represents a
segment of memory space. The stimulus dimensions `MADE OF PAPER', `FLIMSY,
`HAS HOLES' and `CONCAVE' lie on the $x_1$, $x_2$, $x_3$, and $x_4$ axes
respectively.
Three concepts are stored here: `cup', `bag', and `bag with holes'.
Black-ringed dots represent centers of distributed regions where they are
stored. Fuzzy white ball shows region activated by `cup'. Emergence of the
abstract concept `container' implicit in the memory space (central yellow
region) is made possible by the constrained distribution of activation.
\par}

\medskip
\noindent
An extremely general concept such as `depth' is probably even more
nonlocalized. It is latent in mental representations
as dissimilar as `deep swimming pool', `deep-fried zucchini', and `deeply
moving book'; it is deeply woven throughout
the matrix of concepts that constitute one's worldview. In (Gabora 1998,
1999, 2000) one author, inspired by Kauffman's
(1993) autocatalytic scenario for the origin of life, outlines a scenario
for how episodic memories could become collectively
entangled through the emergence of concepts to form a hierarchically
structured worldview. The basic idea goes as
follows. Much as catalysis increases the number of different polymers,
which in turn increases the frequency of
catalysis, reminding events increase concept density by triggering
abstraction-the formation of abstract concepts or
categories such as `tree' or `big'-which in turn increases the frequency of
remindings. And just as catalytic polymers
reach a critical density where some subset of them undergoes a phase
transition to a state where there is a catalytic
pathway to each polymer present, concepts reach a critical density where
some subset of them undergoes a phase
transition to a state where each one is retrievable through a pathway of
remindings events or associations. Finally,
much as autocatalytic closure transforms a set of molecules into an
interconnected and unified living system, conceptual
closure transforms a set of memories into an interconnected and unified
worldview. Episodic memories are now related to
one another through a hierarchical network of increasingly abstract -- and
what for our purposes is more
relevant -- increasingly nonlocalized, concepts.

\subsection{Quantum Structure and the Mind}
Over the past several decades, numerous attempts have been made to forge a
connection between quantum mechanics and the
mind. In these approaches, it is
generally assumed that the only way the two could be connected is through
micro-level quantum events in the brain
exerting macro-level effects on the judgements, decisions, interpretations
of stimuli, and other cognitive functions of
the conscious mind. From the preceding arguments, it should now be clear
that this is not the
only possibility. If quantum structure can
exist at the macro-level, then the process
by which the mind arrives at judgements,
decisions, and stimulus interpretations could itself be quantum in nature.

We should point out that we are not suggesting that the mind is entirely
quantum. Clearly not all concepts and instances
in the mind are entangled or violate Bell inequalities. Our claim is simply
that the mind contains some degree of
quantum structure. In fact, it has been suggested that quantum and classical be
viewed as the extreme ends of a continuum, and
that most of reality may turn out to lie midway in this continuum, and
consist of both quantum and classical aspects in
varying proportions (Aerts 1992; Aerts and Durt 1994).

\section{The Presence of Quantum Probability and Bell
Inequalities}

We have seen that quantum and macroscopic systems can violate Bell
inequalities.  A natural question
that arises is the following: is it possible to construct a
macroscopical  system that violates Bell
inequalities in exactly the same way as a photon singlet state will? Aerts
constructed a very simple model that
does exactly this (Aerts 1991).
This model represents the photon
singlet state before measurement by means of two points that live in the
center of two separate unit
spheres, each one following its own space-time trajectory (in accordance
with the conservation
 of linear and angular momentum), but the two points in the center remain
connected  by means of a rigid but
extendable rod (Fig. 8).  Next the two spheres reach the measurement
apparatuses.
When one side is measured, the
measurement apparatus draws one of the entities  to one of the two possible
outcomes with probability one
half. However, because the rod is between the two entities, the other entity
at the center of the other
sphere is drawn toward the opposite side of the sphere as compared with
the first entity. Only then this
second entity is measured. This is done by attaching a piece of elastic
between the two opposite points of
the sphere that are parallel with the measurement direction chosen by the
experimenter for this side. The
entity falls onto the elastic following the shortest path (i.e.. orthogonal)
and sticks there. Next the
elastic breaks  somewhere and drags the entity towards one of the end
points (Fig. 9). To calculate the probability
of the occurrence of one of the  two possible outcomes of the second
measurement apparatus, we assume there
is a uniform probability of breaking on the  elastic. Next we calculate the
frequency of the coincidence
counts and these turn out to be in exact accordance with the quantum
mechanical prediction.
\noindent

\bigskip
\hskip 3 cm \includegraphics{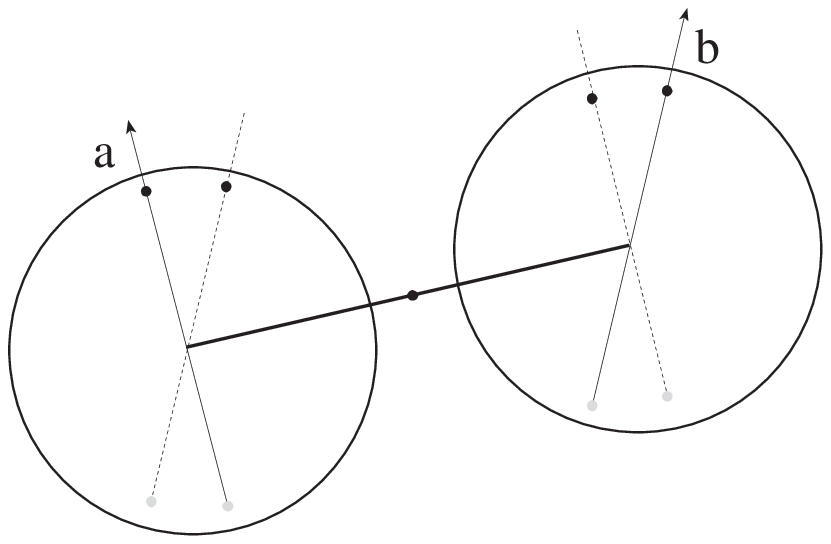}

\bigskip
\noindent {\baselineskip= 7pt \smallroman  Fig\ 8: Symbolic representation
of the
singlet state in the model as two dots in the
centers of their respective spheres.  Also shown is the connecting rod and
the measurement directions chosen at each location a and b.  \par}

\bigskip
\noindent
There are two
ingredients of this model that seem particularly important. First we have
the rigid rod,  which shows the
non-separable wholeness of the singlet coincidence measurement (i.e. the
role of the connecting tube between the vessels of water, or the
associative pathways between concepts). Second, we have the
elastic that breaks which
gives rise to the probabilistic
nature  (the role of the siphon in the vessels of water, or the role of
stimulus input in the mind) of the outcomes.
These two features seem more or less in accordance with
the various opinions researchers have about the meaning of the
violation of Bell inequalities.
Indeed, some have claimed  that the violation of Bell inequalities is due
to the non-local character of the
theory, and hence in our model to the  `rigid but extendable rod', while
others have attributed the
violation not to any form of non-locality, but rather to the  theory being
not `realistic' or to the
intrinsic indeterministic character of quantum theory, and hence to the
role of the  elastic in our model.
As a consequence, researchers working in this field now carefully refer to
the meaning of the
 violation as the ``non-existence of local realism".  Because of this
dichotomy in  interpretation we were
curious what our model had to say on this issue. To explore this we extended
the above model  with
the addition of two parameterizations,
each parameter allowing us to minimize or maximize one of the two
aforementioned features (Aerts D., Aerts S., Coecke B.,
Valckenborgh F., 1995).  The question was of
course, how Bell inequalities would respond to the respective
parameterizations.  We will briefly introduce the model and the results.

\subsection{The Model}

The way the parameterized model works is exactly analogous to the
measurement procedure described above,
with two alterations. First, we impose the restriction that the maximum
distance the rigid rod can
`pull' the second photon out of the center  is equal to some parameter
called $\rho \in [0,1]$. Hence
setting $\rho$ equal to 1 gives us the old situation, while putting $\rho$
equal to zero means we no longer have a
correlation between the two measurements. Second, we allow
the piece of elastic only to break
inside a symmetrical interval $[-\epsilon,\epsilon]$. Setting $\epsilon$
equal to 1 means we restore the
model to the state it was in before parameterization. Setting $\epsilon$
equal to zero
means we have a classical
`deterministic' situation, since the elastic can only break in the center
(there remains the indeterminism of the
classical unstable equilibrium, because indeed the rod can still move in two
ways, up or down).
In fact, to be a bit more precise, we have as a set of states of the entity
the set of couples
\begin{equation}
q \in Q=\{(s_1,s_2)| s_1,s_2,c \in R^3, ||s_1-c|| \leq \rho, ||s_2+c||
\leq \rho , \rho \in [0,1]\}
\end{equation}
Each element of the couple belongs to a
different sphere with center $c$
(resp $-c$ due to linear momentum
conservation)  and whose radius is parameterized by the correlation
parameter $\rho$.  At each side we have a
set of measurements
\begin{equation}
e_1,e_2 \in M= \{ \gamma {\bf n} | {\bf n} \in R^3, ||{\bf n}||=1,
\gamma \in [-\epsilon,+\epsilon],
\epsilon \in [0,1] \}
\end{equation}
The direction $\bf{n}$ can be chosen arbitrarily by
the experimenter at each side
and denotes the direction of the polarizer. (Of course, for the sake of
demonstrating the violation of Bell
inequalities, the experimenter will choose at random between  the specific
angles that maximize the value the
inequality takes).

\noindent
\bigskip
\hskip 3 cm \includegraphics{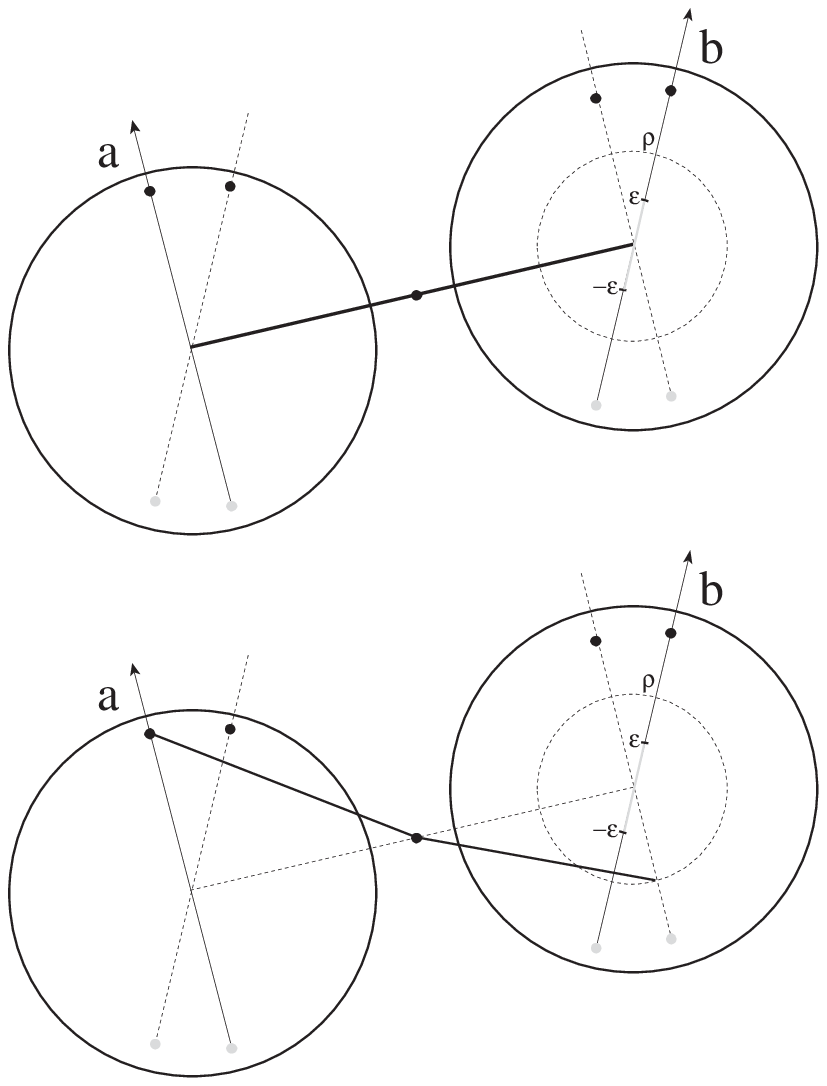}

\noindent {\baselineskip= 7pt \smallroman  Fig\ 9: The situation
immediately before
and after the measurement at one side (left in
this case) has taken place. In the first picture the breakable part of the
elastic is shown (i.e. the interval
$[-\epsilon,+\epsilon]$ on the elastic)  and the maximum
radius $\rho$. In the second picture, we see how the measurement at location a
has altered the state at location b because
of the connecting rod.
\par}

\bigskip
\noindent
The value that the parameter $\gamma$ takes represents
the point of rupture of the
elastic and is unknown in so far that it can take any value inside the
interval $[-\epsilon,+\epsilon]$,
and it will do so with a probability that is uniform over  the entire
interval  $[-\epsilon,+\epsilon]$.
Hence the probability density function related to $\gamma$ is a constant:
$pdf(\gamma)=1/ 2\epsilon$. This represents our lack of knowledge
concerning the specific measurement that
occurred, and is the sole source of indeterminism. As we will show later, if
we take $\gamma$ to vary within
$[-1,+1]$ we have a maximal lack of knowledge about the measurement and we
recover the exact quantum
predictions. As before, the state $q$ of the entity before measurement is
given by the
centers of the two spheres.  The
first measurement $e_1$ projects one center, say $s_1$ orthogonally onto
the elastic that is placed in the
direction $\bf{a}$ chosen by the experimenter at  location A and that can
only break in the interval
$[-\epsilon,+\epsilon]$. Because it will always fall in the middle of the
 elastic, and because we assume a uniform probability of breaking within the
breakable part of the elastic,
the chance that the elastic pulls $s_1$ up is equal to $1/2$ (as is the
probability of it going
down).
Which side $s_1$ is to go depends on the specific point of rupture. Be
that as it may, the experimenter
records the  outcome on his side of the experiment as 'up' or 'down'.  At
the other side of the experiment,
call it location B,  $s_2$ is pulled towards the opposite side of $s_1$
because of the rigid rod connecting
the $s_1$ and $s_2$. The maximum distance it can be pulled away from the
center is equal to $\rho$, and
hence the old $s_2$ transforms to $s_2=-\rho \ \bf{a}$.  Next, experiment
$e_2$ is performed in exactly
the same way as $e_1$.  At location B, the experimenter chooses a
direction, say $\bf{b}$, and attaches the
elastic in this direction.  Again $s_2$ is projected orthogonally onto the
elastic and the elastic breaks.
The main difference between the first and the second measurement is that
$s_2$ is no longer at the center,
but rather at $-\rho \  \bf{a.b}$.

\subsection{Calculating the Probabilities and Coincidence Counts}

There are three qualitatively different parts of the elastic
where $s_2$ can end up after
projection:  the unbreakable part between $[-\rho,-\epsilon]$, the
breakable part between
$[-\epsilon,+\epsilon]$  and the unbreakable part between
$[+\epsilon,+\rho]$. If $s_2$ is in
$[-\rho,-\epsilon]$, then it does not matter where the elastic breaks:
$s_2$ will always be dragged 'up' and
likewise, if $s_2$ is in $[+\epsilon,+\rho]$, the outcome will always be
'down'. If, however, $s_2$ is in
$[-\epsilon,+\epsilon]$, then the probability of $s_2$ being dragged up is
equal to the  probability that
the elastic breaks somewhere in  $[-\rho \ \bf{a.b},-\epsilon]$. This is
the Lebesgue measure of the
interval divided by the total Lebesgue measure of the elastic:
\begin{equation}
P(e_2=up |s_2=-\rho \ {\bf a.b})= { \epsilon -\rho \ {\bf a.b} \over
2\epsilon}
\end{equation}
This settles the
probabilities related to $s_2$.  As before, we define the coincidence
experiment $e_{ij}$ as having four
possible outcomes, namely $(o_i(up), o_j(up))$, $(o_i(up), o_j(down))$,
$(o_i(down), o_j(up))$ and
$(o_i(down), o_j(down))$ (See section 2). Following Bell, we introduce the
expectation values $\EE_{ij},  i,
j \in \{1, 2, 3, 4\}$ for these coincidence experiments, as

\begin{equation}
\begin{array}{ll}
\EE_{ij} = &  +1.P(o_i(up), o_j(up))+1.P(o_i(down), o_j(down))   \\
         & -1.P(o_i(up), o_j(down))-1.P(o_i(down), o_j(up))
\end{array}
\end{equation}

One easily sees that the expectation value related to the coincidence
counts also splits up in three parts.

\begin{itemize}

\item $ -\epsilon < \rho\ {\bf a.b} < +\epsilon$:

In this case we have $P(o_i(up), o_j(up))=P(o_i(down),
o_j(down))=\frac{\epsilon -
\rho\ {\bf a.b} }{4 \epsilon}$ and  $P(o_i(up), o_j(down))=P(o_i(down),
o_j(up))=\frac{\epsilon +
\rho\ {\bf a.b} }{4 \epsilon}$. Hence the expectation value for the
coincidence counts becomes
\begin{equation}
\EE ({\bf a},{\bf b})=- { \rho \ {\bf a.b} \over \epsilon}
\end{equation}
We see that putting $\rho= \epsilon=1$ we get $\EE ({\bf a},{\bf b})=- {\bf
a.b}$, which is precisely the
quantum prediction.

\item $ \rho\ {\bf a.b} \geq +\epsilon$:

In this case we have $P(o_i(up), o_j(up))=P(o_i(down), o_j(down))=0 $ and
$P(o_i(up),
o_j(down))=P(o_i(down), o_j(up))=1/2$. Hence the expectation value for the
coincidence counts becomes
\begin{equation}
\EE ({\bf a},{\bf b})=- 1
\end{equation}
\item $ \rho\ {\bf a.b} \leq -\epsilon$: In this case we have $P(o_i(up),
o_j(up))=P(o_i(down),
o_j(down))=1/2 $ and  $P(o_i(up), o_j(down))=P(o_i(down), o_j(up))=0$.
Hence the expectation value for the
coincidence counts becomes
\begin{equation}
\EE ({\bf a},{\bf b})=+1
\end{equation}
\end{itemize}
\noindent

\subsection{The Violation of Bell Inequalities}

Let us now see what value the left hand side of the Bell inequality takes
for our model for the specific
angles that maximize the inequality. For these angles ${\bf a.b}= \pi/ 4$ and
${\bf a.b}= 3 \pi/ 4$,  the condition   $-\epsilon < \rho\ {\bf a.b} <
\epsilon $ is satisfied only if $
{\sqrt 2 \over 2} < {\epsilon \over \rho}$. In this case, we obtain :
\begin{equation}
\EE({\bf a},{\bf b})=- \EE({\bf a},{\bf b'})=\EE({\bf a'},{\bf
b})=\EE({\bf a'},{\bf b'})= -{\rho \over
\epsilon }\ {\sqrt 2 \over 2 }
\end{equation}
\noindent If, on the other hand we would have chosen $\epsilon$ and
$\rho$ such that $ {\sqrt 2 \over 2} \geq {\epsilon \over
\rho}$ we find:
\begin{equation}
\EE({\bf a},{\bf b})=- \EE({\bf a},{\bf b'})=\EE({\bf a'},{\bf
b})=\EE({\bf a'},{\bf b'})= -1
\end{equation}

\noindent  We can summarize the results of all foregoing calculations in the
following equation:

\begin{equation}
|\EE ({\bf a},{\bf b}) -\EE({\bf a},{\bf b'})|+|\EE({\bf a'},{\bf b})+
\EE({\bf a'},{\bf b'})|  = \left
\{ \begin{array}{cc} \frac{2 \sqrt 2 \rho}{\epsilon} & \frac{\epsilon}{\rho} >
\frac{\sqrt 2}{2} \\ 4 & \frac{\epsilon}{\rho} \leq \frac{\sqrt 2}{2}
\end{array} \right \}
\end{equation}

\noindent For $\epsilon=0$, we have two limiting cases that can easily be
derived: for $\rho=0$ the left
side of the inequality takes the value 0, while for
$\rho \neq  0$  it becomes 4.

For what couples $\rho,\epsilon$ do we violate the inequality? Clearly, we
need only consider the case
${\epsilon \over \rho } > { \sqrt 2 \over 2}$. Demanding that the
inequality be satisfied, we can
summarize our findings in the following simple condition:
\begin{equation}
\epsilon \leq \sqrt2\ \rho
\end{equation}

This result indicates that the model leaves no room for
interpretation as to the source  of the
violation: for any $\rho < { 1 \over \sqrt 2}$, we can restore the
inequality by increasing the amount of
lack of knowledge on the interaction between the measured and the measuring
device, that is by increasing
$\epsilon$. The only way to respect Bell inequalities for all values of
$\epsilon$ is by putting
$\rho=0$.  Likewise, for any $\rho> {1 \over \sqrt 2}$ it becomes
impossible to restore the validity of the
Bell inequalities. The inevitable conclusion is that the correlation is the
source of the violation. The
violation itself should come as no surprise, because we have identified
$\rho$ as the  {\it correlation}
between the two measurements, which is precisely the non-local aspect. This
is also obvious from the fact
that it is this correlation that makes $\EE({\bf a},{\bf b})$ not
representable as an integral of the form
$\int A(a,\lambda)B(b,\lambda)\ d\lambda$ as Bell requires for the
derivation of the inequality. What may
appear surprising however, is the fact that  increasing the indeterminism
(increasing $\epsilon$), decreases
the value the inequality takes! For example, if we take $\epsilon=0$ and
$\rho=1$, we see that the value of
the inequality is 4, which is the largest value the inequality possibly can
have, just as in the case of the
vessels of water model.

\section{Conclusion}
We have presented several arguments to show that Bell
inequalities can be genuinely violated in situations
that do not pertain to the microworld. Of course, this does not decrease
the peculiarity of the quantum mechanical violation
in the EPRB experiment. What it does, is shed light on the possible
underlying mechanisms and provide evidence that the
phenomenon is much more general than has been assumed.

The examples that we have
worked out -- the `vessels of water', the `concepts in the mind', and the
`spheres connected by a rigid
rod' -- each shed new light on the origin of the violation of Bell
inequalities. The
vessels of water and the spheres
connected by a rigid rod examples, show that `non-local connectedness'
plays an essential role in bringing the
violation about. The spheres connected by a rigid rod example shows that
the presence of quantum
uncertainty does not contribute to the violation of the inequality;
on the contrary, increasing quantum uncertainty decreases the violation.

All three examples also reveal
another aspect of reality that plays an important
role in the violation of Bell inequalities: the potential for
different actualizations that
generate the violation. The state of the 20 liters of water
as present in the connected vessels is potentially, but not actually, equal
to `5 liters' plus `15 liters' of water, or
`11 liters' plus `9 liters' of water, {\it etc}. Similarly, we can say that
the concept `cat' is potentially equal to
instances such as our cats, `Glimmer' and `Inkling'. It is this
potentiality that is the
`quantum aspect' in our
nonmicroscopic examples, and that allows for a violation of Bell-type
inequalities.
Indeed, as we know, this potentiality is the fundamental
characteristic of the superposition state as
it appears in quantum mechanics. This means that the aspect of quantum
mechanics that generates the violation of Bell
inequalities, as identified in our examples, is the potential of the
considered state. In the connected vessels example, it is the potential
ways of dividing up 20 liters
of water. In the
concepts in the mind example, it is the
potential instances evoked by the abstract concept `cat'. In the rigid rod
example, it is the possible ways in which the
rod can move around its center.

\section{References}

Aspect, A., Grangier, P. and Roger, G., 1981, ``Experimental tests of
realistic local theories via Bell's
theorem", {\it Phys. Rev. Lett.}, {\bf 47}, 460.

\medskip
\noindent Aspect, A., Grangier, P. and Roger, G., 1982, ``Experimental
realization of
Einstein-Podolsky-Rosen-Bohm gedankenexperiment: a new violation of Bell's
inequalities", {\it Phys. Rev.
Lett.}, {\bf 48}, 91.

\medskip
\noindent Aerts, D., 1981, ``The one and the many", Doctoral dissertation,
Brussels Free University.

\medskip
\noindent Aerts, D., 1982, ``Example of a macroscopical situation that
violates Bell inequalities", {\it
Lett. Nuovo Cim.}, {\bf 34},  107.

\medskip
\noindent Aerts, D., 1984, ``The missing elements of reality in the
description of quantum mechanics of the
EPR paradox situation", {\it Helv. Phys. Acta.}, {\bf 57}, 421.

\medskip
\noindent Aerts, D., 1985a, ``The physical origin of the EPR paradox and
how to violate Bell inequalities by
macroscopical systems", in {\it On the Foundations of modern Physics}, eds.
Lathi, P. and Mittelstaedt, P.,
World Scientific, Singapore, 305.

\medskip
\noindent Aerts, D., 1985b, ``A possible explanation for the probabilities
of quantum mechanics and a
macroscopical situation that violates Bell inequalities",  in {\it Recent
Developments in Quantum Logic},
eds. P. mittelstaedt et al., in Grundlagen der Exacten Naturwissenschaften,
vol.6, Wissenschaftverlag,
Bibliographisches Institut, Mannheim, 235.

\medskip
\noindent Aerts, D., 1991, ``A mechanistic classical laboratory situation
violating the Bell inequalities with
$\sqrt{2}$, exactly 'in the same way' as its violations by the EPR
experiments", Helv. Phys. Acta, {\bf 64}, 1 - 24.

\medskip
\noindent Aerts, D., 1992, ``The construction of reality and its influence
on the understanding of quantum structures",
{\it Int. J. Theor. Phys.}, {\bf 31}, 1815 - 1837.

\medskip
\noindent Aerts, D., 2000, ``The description of joint quantum entities and
the formulation of a paradox",
Int. J. Theor. Phys., {\bf 39}, 483.

\medskip
\noindent Aerts, D., Aerts, S., Coecke, B., Valckenborgh, F., 1995,
``The meaning of the violation of Bell Inequalities:
nonlocal correlation or quantum behavior?", preprint, Free University of
Brussels

\medskip
\noindent Aerts, D. and Durt, T., 1994, ``Quantum. classical and
intermediate, an illustrative example", {\it Found.
Phys. 24}, 1353 - 1368.

\medskip
\noindent
Aerts, D. and Gabora, L., 1999, ``Quantum mind web course lecture week 10"
part of `Consciousness at the Millennium: Quantum
Approaches to Understanding the Mind', an online course offered by
consciousness studies, The University of Arizona,
September 27, 1999 through January 14, 2000.

\medskip
\noindent Aerts, D. and Szabo, L., 1993, ``Is quantum mechanics really a
non Kolmogorovian probability
theory", preprint, CLEA, Brussels Free University.

\medskip
\noindent Bell, J. S., 1964, ``On the Einstein Podolsky Rosen paradox",
{\it Physics}, {\bf 1}, 195.

\medskip
\noindent Bohm, D., 1951, {\it Quantum Theory}, Prentice-Hall, Englewood
Cliffs, New York.

\medskip
\noindent Clauser, J.F., 1976, {\it Phys. Rev. Lett.}, {\bf 36}, 1223.

\medskip
\noindent Clauser, J.F. and Horne, M.A., 1976, Phys. Rev. D, {\bf 10}, 526.

\medskip
\noindent Einstein, A., Podolsky, B. and Rosen, N., 1935, ``Can quantum
mechanical description of physical
reality be considered complete", {\it Phys. Rev.}, {\bf 47}, 777.

\medskip
\noindent Faraci et al., 1974, {\it Lett. Nuovo Cim.}, {\bf 9}, 607.

\medskip
\noindent Freedman, S.J. and Clauser, J.F., 1972, {\it Phys. Rev. Lett.},
{\bf 28}, 938.

\medskip
\noindent
Gabora, L., 1998, ``Autocatalytic closure in a cognitive system: A
tentative scenario
for the origin of culture", {\it Psycoloquy}, {\bf 9}, (67),
[adap-org/9901002].

\medskip
\noindent
Gabora, L., 1999, ``Weaving, bending, patching, mending the fabric of
reality: A cognitive science perspective on
worldview inconsistency", {\it Foundations of Science}, {\bf 3}, (2), 395-428.

\medskip
\noindent
Gabora, L., 2000, ``Conceptual closure: Weaving memories into an
interconnected worldview", in ``Closure: Emergent
Organizations and their Dynamics", eds. Van de Vijver, G. and
Chandler, J., Vol. 901 of the Annals of the New York Academy of Sciences.

\medskip
\noindent
Hancock, P. J. B., Smith, L. S. and Phillips, W. A., 1991, ``A biologically
supported error-correcting learning rule", {\it Neural Computation}, {\bf
3}, (2), 201-212.

\medskip
\noindent
Holden, S.B. and Niranjan, M., 1997, ``Average-case learning curves for
radial basis function networks", {\it Neural
Computation}, {\bf 9}, (2), 441-460.

\medskip
\noindent Holt, R.A. and Pipkin, F.M., 1973, {\it preprint Harvard University}.

\medskip
\noindent
Kauffman, S. A., 1993, {\it Origins of Order}, Oxford University Press,
Oxford, UK.

\medskip
\noindent Kasday, Ullmann and Wu, 1970, {\it Bull. Am.Phys. Soc.}, {\bf
15}, 586.

\medskip
\noindent
Lu, Y.W., Sundararajan, N. and Saratchandran, P., 1997, ``A sequential
learning scheme for function
approximation using minimal radial basis function neural networks", {\it
Neural Computation}, {\bf 9}, (2), 461-478.

\medskip
\noindent Pitowsky, I., 1989, {\it Quantum Probability - Quantum Logic},
Lecture Notes in Physics 321,
Springer, Berlin, New York.

\medskip
\noindent
Willshaw, D. and Dayan, P., 1990, ``Optimal plasticity from matrix
memories: What goes up must come down",
{\it Neural Computation}, {\bf 2}, (1), 85-93.

\end{document}